\documentclass[pdflatex,sn-aps]{sn-jnl}
\usepackage{graphicx} % Required for inserting images
\usepackage{multirow}%
\usepackage{amsmath,amssymb,amsfonts}%
\usepackage{amsthm}%
\usepackage{mathrsfs}%
\usepackage{xcolor}%
\usepackage{textcomp}%
\usepackage{manyfoot}%
\usepackage{booktabs}%
\usepackage{algorithm}%
\usepackage{algorithmicx}%
\usepackage{algpseudocode}%
\usepackage[utf8]{inputenc}
\usepackage[normalem]{ulem}
\usepackage{listings}\usepackage{rotating} % Rotating table
  \usepackage{tcolorbox}
\usepackage[
singlelinecheck=false % <-- important
]{caption}
\usepackage{etoolbox}   
\makeatletter
\patchcmd{\@maketitle}{\artauthors}{\centering{\artauthors}}{}{}
\makeatother

\newcommand{\blue}[1]{\emph{\color{blue} #1}}

\begin{document}
\title{\vskip -0cm \hskip 0.3cm First comments on a descoped/staged FCC-ee}
%\title{\vskip -2cm \hskip 0.7cm Comments on the CERN flagship collider \\ preferred and preferred alternative options}
\subtitle{\vskip -0.2cm \large Contribution to the 2026 European Strategy for Particle Physics\\
}
%\date{21 June 2023}

\author[1]{\fnm{A.} \sur{Blondel}}
\author[2,3]{\fnm{C.} \sur{Grojean}}
\author*[4]{\fnm{P.} \sur{Janot}}\email{patrick.janot@cern.ch}
\author[5]{\fnm{G.} \sur{Wilkinson}}

\affil[1]{\small \orgname{LPNHE Paris-Sorbonne}, \orgaddress{\street{4 place Jussieu}, \city{Paris}, \postcode{75252}, \country{France}}}

\affil[2]{\small \orgname{Deutsches Elektronen-Synchrotron DESY}, \orgaddress{\street{Notkestr. 85}, \postcode{22607} \city{Hamburg}, \country{Germany}}}

\affil[3]{\small \orgname{\orgdiv{Institut für Physik}, Humboldt-Universit\"at zu Berlin}, \orgaddress{\postcode{12489} \city{Berlin}, \country{Germany}}}

\affil[4]{\small \orgname{CERN}, \orgdiv{EP Department}, \orgaddress{\street{1 Esplanade des Particules}, \city{Meyrin}, \country{Switzerland}}}

\affil[5]{\small \orgname{University of Oxford}, \orgdiv{Department of Physics}, \orgaddress{\city{Oxford}, \country{United Kingdom}}}

\abstract
{In response to its remit, the European Strategy Group (ESG) recommended the electron–positron Future Circular Collider (FCC-ee)
as the preferred option for the next flagship collider at CERN; and a descoped FCC-ee as the preferred alternative option (with reduced synchrotron radiation (SR) power, without a run at the $\rm t\bar t$ threshold, and with only two interaction regions) in the event that the preferred option turns out not to be feasible. 

Upon request of the ESG, a basic comparison of the physics potential of the descoped option with that of the baseline version is presented in this short note. Our first observations about the alternative proposal of a descoped FCC-ee are that {\it (i)} the same performance as the baseline, apart from $\rm t \bar t$-run related, is achieved with the longer proposed period of operation; and {\it (ii)} the top run remains essential, but can be staged at a later date. The full appraisal of the consequences of descoping and a proposal for the best way to integrate possible staging wherever feasible, and possible improvements, will need further joint studies by the physics and accelerator groups.

}
\maketitle

%\tableofcontents

%\vfill\eject 

\section{Introduction}

According to the Council remit~\cite{ESGRemit} for the European Strategy Group (ESG), {\it the aim of the Strategy update should be to develop a visionary and concrete plan that greatly advances human knowledge in fundamental physics through the realisation of the next flagship project at CERN. This plan should attract and value international collaboration and should allow Europe to continue to play a leading role in the field.}

After studying, for the various flagship project proposals, the discovery potential; the competitiveness; the financial feasibility; the technical readiness; the possibility of a viable path towards partonic centre-of-mass energy well in excess of 10\,TeV; and the community input; the following recommendations were drafted on 5 December 2025 and made public on 12 December 2025~\cite{CERN-ESU-2025-002} at the end of the December Council meeting. 

\noindent \blue{A. The electron–positron Future Circular Collider (FCC-ee) is recommended as the preferred option for the next flagship collider at CERN.}

{\it The FCC-ee would deliver the world’s broadest high-precision particle physics programme, with an outstanding discovery potential through the Higgs, electroweak, flavour and top-quark sectors, as well as advances in quantum chromodynamics (QCD). Its technical feasibility is demonstrated by the comprehensive FCC Feasibility Study, its scope and cost are well defined, plausible funding scenarios have been developed and its schedule enables first beams within five to seven years after the end of HL-LHC operations. The FCC-ee would maintain European leadership in high-energy particle physics, as well as advancing technology and providing significant societal benefits. The FCC-ee would also pave the way towards a hadron collider reusing the tunnel and much of the infrastructure, providing direct discovery reach well beyond the 10\,TeV parton energy scale, in line with the community’s ambition for exploration at the highest achievable energy. The overwhelming endorsement of the FCC-ee by the particle physics communities of CERN’s Member and Associate
Member States further reinforces it as the preferred path.}

\noindent \blue{B. A descoped FCC-ee is the preferred alternative option for the next flagship collider at CERN.}

{\it Descoping scenarios include removing the top-quark run, constructing two rather than four interaction regions and experiments and decreasing the radiofrequency (RF) system power. These measures would reduce the construction cost by approximately 15\%. Although this would have a significant impact on the breadth of the physics programme and the precision achieved, the descoped FCC-ee would still provide a very strong physics programme and a viable path towards high energies, compared to the alternative collider options. Should additional resources become available, these descoping scenarios would be reversible.}

\begin{figure}[htbp]
\centering
\includegraphics[width=0.95\textwidth]{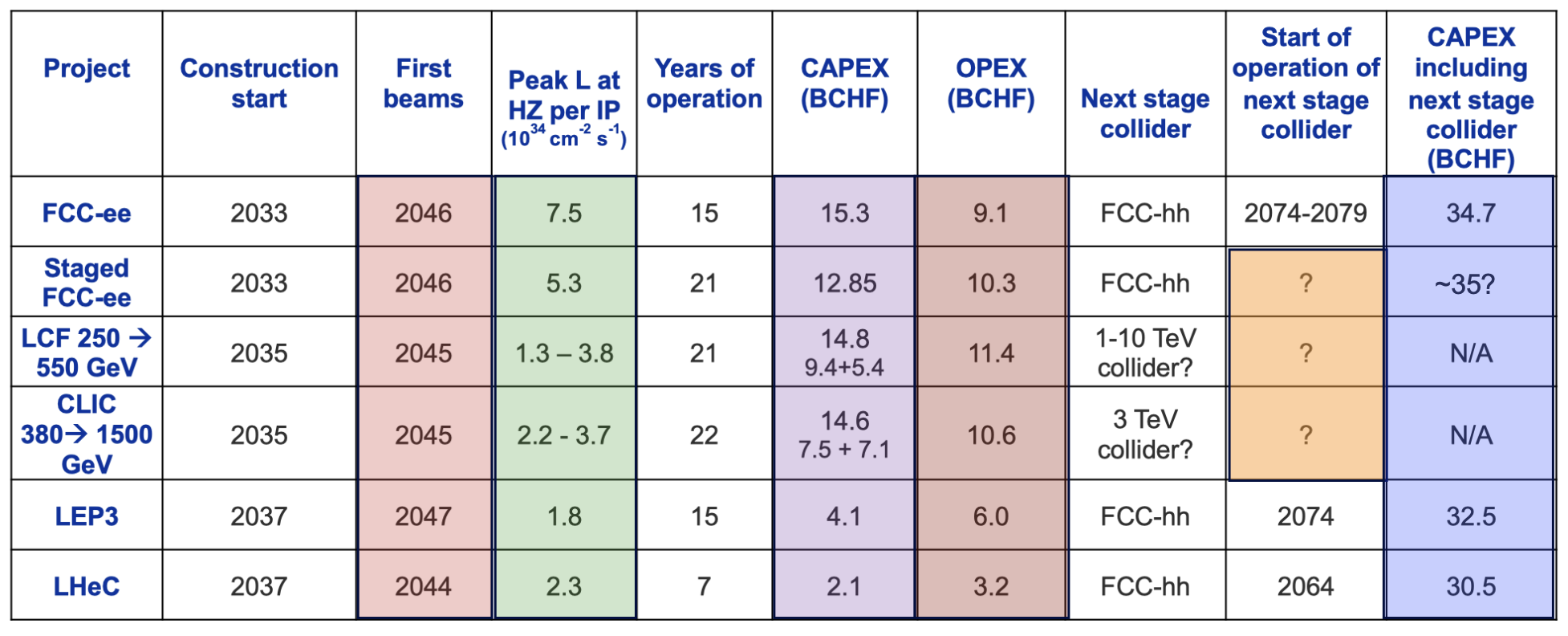}
\caption{\label{fig:DescopedDetails} \small From slide 100 of Ref.~\cite{Fabiola2025}: details about the descoped FCC-ee (dubbed ``staged FCC-ee'' in the table) and the other colliders discussed in the ESG drafting session, in particular: instantaneous luminosity per IP for ZH productions, duration of operation, capital cost, operation cost, path towards higher energies.}
\end{figure}

More details were given on the concrete scenarios for a descoped FCC-ee in a subsequent presentation~\cite{Fabiola2025} from the CERN Director General to the CERN personnel on 16 December 2025. In addition to the possible removal of the runs at the top-pair production threshold and above ($\sqrt{s} > 340$\,GeV) and/or the construction of only two interaction points (IPs) instead of four, the synchrotron radiation (SR) power would be reduced from 50 to 30\,MW per beam -- complemented by a commensurate descoping of the technical infrastructure (reduced thermal load for cooling and ventilation, resized electrical distribution, simplified cryo-modules, halved injector repetition rate). The corresponding savings were estimated to amount to 1.26, 0.80 and 0.35 billion Swiss Francs (CHF), respectively. Additional information is displayed in Table~\ref{fig:DescopedDetails} (from slide 100 of Ref.~\cite{Fabiola2025}). Of particular interest %are the instantaneous luminosities and the operation time of all flagship project proposals: to partly compensate for the loss of instantaneous luminosity per IP at a descoped FCC-ee (due to the SR power reduction), the operation time is increased from 15 to 21 years with respect to the preferred baseline FCC-ee.  
is the $\sim 21$ years of operation for the descoped FCC-ee (dubbed ``staged FCC-ee'' in the table) needed, within $\pm 2$ years, to absorb the cumulative budget deficit (CBD) created by the FCC-ee construction~\cite{FabiolaPrivate}.   

On 18 December 2025, the ESG Working Group 2b (WG2b) -- in charge of the assessment of the overall physics potential of the different options -- contacted us to understand how the statistical and systematic uncertainties would scale for the precision measurements at a descoped FCC-ee. We now try to give a preliminary answer to this request. %in Section~\ref{sec:DescopedPrecision}. %The relevance of each of the descoping/staging items is then discussed in Section~\ref{sec:discussion}. Our observations and recommendations are summarised in Section~\ref{sec:summary}.

\section{Scaling measurement uncertainties at a descoped FCC-ee}
%\label{sec:DescopedPrecision}

 The instantaneous luminosity ${\cal L}$ is expected to scale linearly with the SR power: a decrease of the latter from 50 to 30\,MW therefore reduces ${\cal L}$ by a factor 0.60. The effect of the removal of two IPs is more difficult to estimate: as soon as the FCC ring layout was re-optimised (following the first high-level objective set by the Council for the feasibility study in 2021~\cite{FCC-FS-plans:202106}) and made compatible with four IPs thanks to its new fourfold super-periodicity, the collider optics was designed only under the hypothesis of four interaction regions. This choice was supported in the recommendations of all FCC mid-term review committees in 2023, so that no optics with only two IPs exists today. At the end of the FCC Conceptual Design Study, however, a rough estimate had been made that going from two to four IPs would reduce the instantaneous luminosity by approximately 15--20\% at each interaction region. In the absence of dedicated studies, the same estimate would induce a reduction of the total instantaneous luminosity ${\cal L}$ (i.e., summed over all IPs) by another factor $0.61 \pm 0.02$ when removing two collision points. 
 
 When both effects are combined, the total instantaneous luminosity of the descoped FCC-ee (summed over the two IPs) would therefore be smaller than that of the baseline FCC-ee (summed over the four IPs) by a factor $0.365 \pm 0.010$.  The baseline FCC-ee operations include four years at and around the Z pole (beginning with an initial two years at half the luminosity), two years at the WW threshold, three years at the ZH rate maximum, and five years at and above the $\rm t \bar t$ threshold. To get the same integrated luminosity (as shown in Table~\ref{tab:seqdescoped}) for the first three energy stages, and therefore the same statistical uncertainties, the descoped FCC-ee would require running $9\frac{1}{4}$ years at and around the Z pole (of which two with half the luminosity), $5\frac{1}{2}$ years at the WW threshold, and $8\frac{1}{4}$ years for ZH production, for a total of 23 years. In first approximation, this duration is the same as the $21\pm 2$ years proposed in Table~\ref{fig:DescopedDetails}. 
 
\begin{table}[ht]
\renewcommand{\arraystretch}{1.10}
\centering
\caption{\small A descoped/staged FCC-ee operation model with two IPs and a SR power reduced to 30\,MW, designed to reach the same integrated luminosity as the baseline FCC-ee. The five rows show the centre-of-mass energies, the descoped instantaneous luminosities for each IP, the integrated luminosities per year summed over 2~IPs, the run durations, and the total integrated luminosities. The integrated luminosities delivered during the first two years at the Z pole and the first year at the $\rm t \bar t$ threshold are half of the annual design value.} 
\label{tab:seqdescoped}
\begin{tabular}{lccccc}
\hline 
Working point & Z pole & WW thresh.\ & ZH & \multicolumn{2}{c}{(Staged) $\rm t\bar t $} \\ \hline
$\sqrt{s}$ {(GeV)} & 88, 91, 94 & 157, 163 & 240 & 340--350 & 365 \\ 
Lumi/IP {($10^{34}$\,cm$^{-2}$s$^{-1}$)} & 105 & 14.6 & 5.46 & 1.31 & 1.02 \\ 
Lumi/year {(ab$^{-1}$)} & 25.2 & 3.5 & 1.3 & 0.31 & 0.24 \\ 
Run time {(year)} & $9\frac{1}{4}$ & $5\frac{1}{2}$ & $8\frac{1}{4}$ & $2$ & $11$ \\ 
Integrated lumi.\ {(ab$^{-1}$)} & 207 & 19.2 & 10.8 & 0.47 & 2.69 \\ \hline
\end{tabular}
\end{table}

As discussed in Ref.~\cite{selvaggi_2025_n78xk-qcv56}, most experimental systematic uncertainties of FCC-ee precision measurements scale with the integrated luminosity. With the proposed descoped FCC-ee in 23 years, all statistical uncertainties and most systematic uncertainties would therefore be identical to those achieved with the baseline FCC-ee in nine years.\footnote{We note in passing that an additional SR power of 3\,MW per beam, representing an additional cost of 50 million CHF (while still reducing the total construction cost by slightly more than 15\% together with the other two descoping items), would suffice to get exactly the same uncertainties in 21 years as the baseline FCC-ee would in nine years. This small additional investment would be a more cost-efficient way to recover the full precision of the baseline FCC-ee than the extra two years of operation that would be needed with the 30 MW machine, saving about twenty times more money ($\sim 1$\,BCHF).}  

% Importantly, some systematic uncertainties do not change with a reduction (or an increase, for that matter) of the integrated luminosity. Theoretical uncertainties are of this kind, even if smaller statistical uncertainties would certainly be a strong motivation to further improve the theoretical precision: nobody wants to be responsible for missing a discovery at FCC-ee! 

The control of some systematic biases, however, may depend on the number of detectors, even with the same integrated luminosity. Maintaining the same accuracy on the set of electroweak precision observables is generally possible, given that systematic uncertainties are based on (or controlled by) measurements obtained in the same dataset. In some cases, however, it may create further challenges.  For example, the alignment requirements on the low angle fiducial cut, such as spelled out in Ref.~\cite{blondel_2023_f1fs5-0jr59}, would need to be tightened by a factor $\sqrt{2}$ (from typically 10 to 7\,$\mu$m  at a distance of 2.5\,m) to keep the same level of accuracy for the luminosity measurement with $\rm e^+e^- \to \gamma\gamma$ events or for the ratio of hadronic to leptonic decays of the Z. These additional difficulties may be largely alleviated by the in-situ acceptance measurement proposed in Ref.~\cite{JanotFCCWeek}.
 
 The precision of the determination of the average circulating beam energy with resonant depolarisation (RDP), at the Z pole and the WW threshold, is independent on the configuration of the machine. Additional effects due in particular to energy losses both in collisions (beamstrahlung) and around the arcs ({\it e.g.}, energy losses in quadrupoles) depend on the specific configuration, and must be determined experimentally. One particularly sensitive way to monitor the energy losses is to measure the boost at each of the IPs using muon pairs. This measurement is dependent on the integrated luminosity, on the number of IPs, and on their arrangement. It is definitely better done with four experiments: with only two experiments, it is not possible to isolate the region of the machine where an unexpected loss would occur. The level of systematic precision presented in the ESPPU2025 inputs~\cite{selvaggi_2025_n78xk-qcv56, blondel_2025_5madt-cfy16} -- 100\,keV for the Z mass, 12\,keV for the Z width, 150\,keV for the W mass -- is however quite conservative and would therefore not change significantly with only two experiments. Progress made since March 2025~\cite{YiWu20251014} shows that significantly improved precision may be attainable on the knowledge of the mean beam energy. In this case, it would be essential to have information available from all four IPs in order to benefit fully in the determination of the collision energy. %Interestingly, progress made on this front since March 2025 shows that the RDP can be performed with a precision on beam energy of the order of 5\,keV~\cite{YiWu20251014}. This breakthrough should lead to much improved numbers, for which the availability of four IPs will be essential.

 {\bf In summary, we therefore suggest that the ESG WG2b should keep exactly the same uncertainties as in the briefing book for the descoped FCC-ee version}, albeit without the $\rm t \bar t$ run. 

\section{A staged top run}
%\label{sec:discussion}

 %With the above suggestion to use the same uncertainties for the descoped FCC-ee and the baseline FCC-ee, are we not then saying that the baseline FCC-ee is over-designed, and maybe that an even more descoped FCC-ee version would be advantageous? As explained in detail in Volume 1 of the Feasibility Study Report (FSR)~\cite{FCC:2025lpp}, the short answer to these questions is obviously no. The main arguments to build the baseline FCC-ee rather than any descoped version are recalled below. 
 
%\subsection{Why is a $\rm t \bar t$ run of great scientific relevance?}
%\label{sec:ttbar}
%(To accumulate the same integrated luminosity at the baseline FCC-ee, a staged top run would have to last 13 years, of which two for the mass measurement, the first one with half the luminosity.)

Now, even with the same measurement uncertainties up to 240\,GeV, the absence of a run at higher energy would still degrade the interpretations of the low-energy measurements, which makes a $\rm t\bar t $ run of great relevance. 

First, a short run to scan the $\rm t \bar t$ threshold ($\sqrt{s} = 340$--350\,GeV) allows the measurement of the top-quark mass, a fundamental parameter of the Standard Model, with a precision of $\cal O$(10\,MeV)~\cite{Defranchis:2025auz}, for which hadron colliders cannot compete. Because the prediction of electroweak precision observables (EWPO) is, in various ways, sensitive to $m_{\rm top}$, the discovery power of this EWPO exploration is limited by the uncertainty on $m_{\rm top}$. For example, matching the precision of the SM predictions from the EWPO measurements to the 180\,keV (resp.\ 4\,keV) statistical uncertainty on the W mass (resp.\ Z width) requires a 20\,MeV (resp.\ 15\,MeV) knowledge of $m_\text{top}$.  It is only when the mass of the top quark is measured with this precision that the FCC-ee EWPO measurements give their best sensitivity, typically increasing the reach in the new physics energy scale by 60\%. 

Also, the ZH cross section dependence on $\sqrt{s}$ provides sensitivity to the Higgs boson self-coupling when data at 365\,GeV are available, allowing a standalone 5\,$\sigma$ discovery of this coupling to be contemplated at the baseline FCC-ee (or with $11+2$ years at a descoped/staged FCC-ee top run). Reducing the descoped/staged top run duration to $4+2$ years would still allow a $5\sigma$ discovery, in combination with HL-LHC, but the Higgs coupling precision in a global SMEFT fit would deteriorate. More importantly, the per-cent measurement of the top EW couplings  {\it(i)}~matches the EWPO ppm precision at the Z pole and the WW threshold;  and {\it (ii)}~keeps the theoretical uncertainties on the top Yukawa coupling determination at FCC-hh at the per-cent level, a pre-requisite for the model-independent determination of the Higgs boson self-coupling at FCC-hh.

Altogether, a FCC-ee run at the top-pair production threshold and above allows the optimal use of both the FCC-ee data at lower energies and the FCC-hh data, in their interpretation in terms of new physics. A descoped FCC-ee without a $\rm t \bar t$ run being, at least, programmed at a later stage, would seriously affect the precision measurement programme and would substantially reduce the discovery power of FCC. 

{\bf The top run should thus not be suppressed but,  instead, staged}.\footnote{As a matter of fact, the top run has always been considered, since the beginning of the FCC conceptual design study in 2014, as a second stage of the FCC-ee programme.} %Everything must be done to find the 1.26\,BCHF needed to reach this centre-of-mass energy at FCC-ee, dominated by the construction of the 800\,MHz 9\,GV RF system. The good part is that this amount is not needed 

The additional amount of 1.26 BCHF, dominated by the cost of the 800 MHz 9\,GV RF system,  need not be committed at the beginning of the construction of FCC-ee, as the top run would occur 23 years (or 10 years for the baseline FCC-ee) after the start of operations. At that point, not only would more money become available due to the resorption of the CBD,  but this long period at lower energy would also open the possibility of an in-kind contribution delivered just in time for the one-year shutdown prior to the highest energy run. 

{\bf To allow for this essential upgrade, the machine should be designed from the beginning to withstand the running conditions at the top-quark pair production energies.}

\section {Summary and outloook}
To summarise, it has been shown/recalled that {\it (i)} the measurements achieved at a descoped FCC-ee (only 2 IPs, 30\,MW of SR power per beam, and no top run)  can be made as precise as those from the baseline FCC-ee without a top run, at the expense of a longer operation time, as anticipated by the ESG in the definition of the descoped FCC-ee; and {\it (ii)} a staged $\rm t \bar t$ run, scheduled at the end of the descoped FCC-ee operation (thus giving time to identify the relevant funding sources), very substantially expands the discovery power of both FCC-ee and FCC-hh. 

We therefore suggest for the short term that the ESG WG2b should keep exactly the same uncertainties as in the briefing book for the descoped FCC-ee, either without a $\rm t\bar t$ run or, better, with a staged top run after 23 years, when the cumulative budget deficit allows for it. Should financial reasons prevent the construction of the entire baseline FCC-ee project, we recommend that a descoped FCC-ee be designed from the beginning to allow for the essential top run upgrade at a later stage. 

We also would like to recall that the possibility of running four interaction points is a feature unique to circular colliders. 
It is remarkable that the interaction points themselves present an excellent opportunity for very large private contributions to be recognised through the naming of individual experimental sites in a manner that acknowledges the individual donors. On 18 December 2025, {\it i.e.}, after the ESG recommendations were issued, a press release announced that already three private donors have pledged significant funds towards the construction of the Future Circular Collider (FCC)~\cite{pressRelease}. These contributions, totalling some 860 million euros, would already suffice to consolidate the four interaction points of FCC-ee. 

Finally, we note that the synchrotron radiation power per beam at the baseline FCC-ee is the same 50\,MW as that of its predecessors (LEP3, DLEP~\cite{blondel2012highluminosityeecollider} in 2011, and TLEP~\cite{Blondel:2012ey} in 2012). It was not chosen as a result of an optimisation, but followed a recommendation by the CERN Director of Accelerators and Technology back in 2009 for the electron branch of LHeC. This arbitrary value was never changed since. The recent recommendation of the ESG to reduce this SR power to yet another arbitrary value (30\,MW) for the descoped FCC-ee is a good incentive to review the relative merits of the different choices. The optimisation of the SR power with respect to performance and cost is, among other important things, the topic of ongoing work, and will be reported soon enough. 

%and {\it (iv)} running FCC-ee with a SR power upgraded to 100\,MW (instead of 50), four interaction points and a top run might save up to 3.8\,BCHF with respect to the proposed descoped FCC-ee, and up to 3.0\,BCHF with respect to the current baseline FCC-ee. 

%In conclusion,  and strongly advocate that the \underline{irreversible} decision of going to 2 IPs be absolutely avoided. Finally we argue  that an improved baseline FCC-ee, with four interaction points, a $\rm t\bar t$ run, and 100\,MW synchrotron radiation power per beam (instead of 50\,MW per beam), offers the best scientific and financial prospects. We consequently advise that a financial and technical feasibility study of such an RF power upgrade (and any other means to increase the specific luminosity, such as a reduction of the vertical $\beta^\ast$) should be undertaken without further ado.  

\section*{Acknowledgements}

 This work has been partially funded from the European Union's Horizon 2020 research and innovation programme under grant agreement No. 951754, and is supported by the Deutsche Forschungsgemeinschaft (DFG) under Germany’s Excellence Strategy EXC 2121 ``Quantum Universe'' -- 390833306, as well as by the DFG grant 491245950. %This project also has received funding from the European Union’s Horizon Europe research and innovation programme under the Marie Sk\l{}odowska-Curie Staff Exchange grant agreement No 101086085 - ASYMMETRY. 
We are indebted to Carlos Lourenço and Fabiola Gianotti for their comments to the manuscript. 
%\appendix
%\appendixpage
%\addappheadtotoc

\bibliography{References}
\end{document}